\documentclass[12pt]{article}
\usepackage{amsfonts}
\usepackage{amssymb}
\usepackage{amsmath}
\renewcommand{\d}{\delta}
\renewcommand{\b}{\beta}
\newcommand{\g}{\gamma}

\newcommand{\ar}{\longrightarrow}

\renewcommand{\a}{\alpha}
\begin{document}
\title{Dynamical diffusion as the approximation of one quantum particle dynamics}
\author{Y.Ozhigov\thanks{Work is supported by NIX Computer Company, grant \# F793/8-05, INTAS grant 04-77-7289, and Russian Fond for Basic Researches, grant 06-01-00494-a. e-mail: ozhigov@cs.msu.su} \\[7mm]
Moscow State University,\\
Institute of physics and technology RAS 
} 
\maketitle

\begin{abstract}
The paper contains the proof that the diffusion ensemble of point wise particles with the intensity depending on the grain of spatial resolution serves as the satisfactory approximation of one quantum particle dynamics. 
\end{abstract}

\section{Statement of problem}

The work continues the algorithmic approach to quantum mechanics that has been initiated in the works \cite{Oz1}, \cite{OAO}. The main aim of algorithmic approach is to build the video film reflecting the dynamics of many quantum particles. We believe that the modern computers make possible to create such video films that reflect the dynamics of billions quantum particles including all known quantum effects. Such video films must show all details of process that could be obtained by any sequence of observation fulfilled on the considered system. 
   The creation of such picture we call the simulation of quantum dynamics. The simulation can be based on quantum theory only, because it is the single theory of micro world. At the same time the simulation of quantum dynamics represents the serious difficulty due to the fundamental difficulty of quantum computations. This difficulty lies in the formalism of quantum theory: the dimensionality of the space of states grows exponentially when the total number of particles grows. 

Though, the computational difficulties arise already for one quantum particle. Let its configuration space be divided to $N$ elements. It means that the space of quantum states has the dimensionality $N$. If we apply the matrix algebra in any form for computations we force the computer to process all trajectories of the system passing through all $N$ basic states. In the simplest case it is expressed in the multiplication of unitary matrices of the time evolution. The mean value of the module of matrix element is $1/\sqrt{N}$. If all the interference arising in the evolution of the system is constructive we would obtain the resulting matrix which coefficients are about $N\ 1/(\sqrt{N}\sqrt{N}=1$, whereas they must be of the order $1/\sqrt{N}$. It means that the bulk of interference is destructive and the huge portion of the computational recourse is spent to verify only that there is no particle in the considered point. The computational methods of matrix algebra when applied to quantum mechanics a priori require the no efficient expenses of the computational resources. Such methods factually realize Ryman scheme of integration for Schr?dinger equation that is based on the division of the configuration space to finite elements. 

We propose the alternative approach corresponding to Lebeg scheme of integration. We take as the basic elements the point wise samples of the considered particle so that each sample will represent the whole particle in one point. The total number of samples must not be large and its dynamics must be the good approximation of the quantum dynamics of the initial particle. Our main aim is to avoid the huge non effective expenses of the computational resources featured to matrix algebra. After the developing of this technique for one quantum particle we could hope to apply this method for many quantum particles as well. 

This method of representation of quantum particle by its samples we call the method of collective behavior. It realizes the requirement of strict economy of the computational resources, which is the basic principle of algorithmic approach. The requirement of maximal economy of the computational resources is not only esthetic. This requirement allows to build the models in which decoherence is inbuilt feature of the model but not an axiom as in the Copenhagen quantum mechanics. In is shown in \cite{Oz1} how this requirement gives us the classical urn scheme for Born distribution of probability for the results of quantum measurements.  

Here we give the interpretation of the dynamics of one quantum particle by means of collective behavior. The cost that will be paid for the economy of the computational resources is the necessity to build the algorithm by means of mechanism of interaction between the samples of particle, not by the differential equations. For the method of collective behavior it is impossible to build the adequate differential equation. However, this situation has the positive sides as well, beyond economical computations. The model of quantum dynamics becomes nearer to classical than in the standard approach that makes possible its visual representation. 

The proposed approach is the direct generalization of diffusion Monte Carlo method to the case of the time dependent solution of Schr?dinger equation. The known fact that DMC gives the most exact approximation of stationary wave functions among all computational methods inspires optimism in the practical application of the method of collective behavior for more complex problems.

\section{Dynamical diffusion swarm}

   Here we define the main instrument of quantum simulation: dynamical diffusion swarm. This object generalizes two well known notions: the ensemble of point wise particles from DMC method, and the ensemble of particles with the interaction induced by some classical Hamiltonian $H(r,p)$. Particles from DMC have no speeds and they are designed for the computation of stationary states for which they give the best approximation. The density of particles $\rho$ for the ensembles with classical Hamiltonian depends on the coordinate $r$ and on the impulse $p$; it obeys Liouvill equation 
$$
\frac{d\rho}{dt}=-\{ \rho, H \} .
$$
The behavior determined by this equation cannot simulate quantum evolution with the admissible accuracy because it does not give principle quantum phenomena like Rabi oscillation or quantum spectra. Hence, for the simulation of quantum mechanics we must admit some elements of behavior of the samples which do not follow from the classical physics. 

The next analog of the dynamical diffusion swarm is Calder Legett model for the partial decoherence of quantum particle in which the particle is considered as interacting with the bath of harmonic oscillators. This interaction gives some random speed to the particle. However, such a model is based on standard formalism whereas the dynamical diffusion swarm is designed to replace this formalism.

 Why the dynamical diffusion swarm is better than the immediate solution of Shroedinger equation ? In this solution we factually use Riemann schem of integration. We must perform computations of the wave function on the whole configuration space independently on the degree of constructiveness of the interference. Here on the main part of the space where the interference is destructive and the wave function factually equals zero we are forced to spend the computational recourse only to verify this. The dynamical diffusion swarm, in contrast, realizes the more general Lebeg scheme og integration. In this case the diffusion dynamics results in that the samples will concentrate in the areas of constructive interference themselves and we avoid the non effective expenses of the computational resources. This is the fundamental advantage of the diffusion dynamics. We will see that the cost for this is the non uniform dependence of diffusion rate on the grain of the length $\d x$ in contrast to the standard diffusion where the rate is uniform. 

   We proceed with the definitions. We call the swarm the finite set $S$ consisting $n$ identical point wise particles each of which $s\in S$ has its own coordinates and impulse $x(s),\ p(s)\in R^3$. In the method of collective behavior one quantum particle of the mass $M$ and charge $Q$ is represented by the swarm $S$ each member $s\in S$ of which has the mass $m=M/n$ and the charge $q=Q/n$. The elements of this swarm are called the samples of this quantum particle. We suppose that the total number of samples $n$ is so large that the swarm can be used as the approximation of the continuous media. E.g., if we need use the smaller and smaller spatial grain some samples will always occur in each spatial cell. But the dispersion of speeds will grow when the grain decreases, and we will have the separate swarm for each spatial grain $\d x$. 

   The choice of spatial grain is closely connected with by the definition what object must be considered as quantum particle. Thiis definition in turn depends on the concrete problem and quantum particles are not necessary elementary in the sense of theoretical physics. The definition what must be treated as a particle presumes the choice of the typical length $\Delta X$ and the time $\Delta T$, so that the size of particle is much lesser than  $\Delta X$, e.g. it can be treated as point wise, and the time interval $\Delta T$ is not less than the typical time of the processes we are interested in. Let us agree that the typical mean speeds of the considered shifts are much lesser than some limit speed of all movements $c$. For example, an atom can be treated as a point wise particle in he processes with $\Delta X > 10^{-8}m$ and $\Delta T>10^{-10}s$. If we decrease the value of typical lengths and times then to obtain the right picture we must consider the different set of elementary particles, for example, the separate nucleus and electrons inside of atom. If we fix $\Delta X$ and $\Delta T$, then to obtain the dynamical picture we must define  the smaller segments $\d x$, $\d t$, which will represent the elementary steps of the video film and which, however, must be much greater than the typical lengths and times $\tilde\Delta X,\ \tilde\Delta T$ of the more fundamental processes then the considered one (the gap between the different fundamental processes can be about $10^{-20}$, that always allows to make this separation). Also in the same process with the fixed energy the times and lengths depends on masses. The separation of particles by their masses makes possible to consider for the bulk of processes in electrodynamics only electrons because the typical distances of flight of protons are to $1800$ times smaller. We can then treat the chosen values $\Delta X$, $\Delta T$ as the size of imaginary screen and the length of video film, and $\d x$, $\d t$ as the grain of spatial resolution of screen and the time of showing of one card in the film. We choose $\d x$ and $\d t$ maximal so that our film will be informative. After this choice the conclusion can be done about what particle should be treated as quantum and what as classical. For this the typical values of their action $a=M(\Delta X)^2/\Delta T$ should be compared with Plank constant $h$. If $a<h$ then the particle should be treated as quantum, otherwise as classical. In the method of collective behavior the passage from classical to quantum type of consideration means simply the change of swarm size, e.g. does not mean the different type of dynamics. Due to the above mentioned reserve in the choice of resolution in the process of film preparation we then can further decrease the values $\d x$ and $\d t$ for the forming of right picture , for example, dividing these intervals to smaller parts and obtain the better approximation to the solution of Shroedinger equation. We assume that the space $R^3$ is divided to the equal cubes with the side $\d x$, and the time is divided to the equal intervals of the longitude $\d t$. 

We introduce the value $c$, which is the single nonzero speed of movement of the samples in the swarm. The intervals of time and distances will be always chosen so that $\d x \gg c\d t$. It guarantees that in each step of the evolution the values of magnitudes obtained as the mean values on cubes with the side $\d x$ will change small that is necessary for the asymptotical approximation. 

The density of swarm in the point $x$ is determined by the expression
\begin{equation}
\rho (r, t) = \frac{N(r,t)}{(\d x)^3},
\label{density}
\end{equation}
where $N(r,t)$ denotes the total number of samples occurring in the moment $t$ in the same cube with the point $r$. For the comparison with Shroedinger equation in this definition we should converge $\d x \ar 0$, that means the consideration of the sequence of the swarms with the densities $\rho_n$ with the growing $n$ instead of one swarm. Further we assume that the value $\d x$ is fixed. We write $\rho(x)=|\Psi(x)|^2$ instead of 
\begin{equation}
\rho_n(x)\ar |\Psi(x)|^2 (n\ar \infty ),
\label{asympt}
\end{equation}
where the convergence is uniform without special mentioning. This sequence of the swarms realizing the approximation of the exact solution of Shroedinger equation is called the admissible approximation of quantum evolution. 
    
    Our aim is to define the behavior of the samples in the swarm which gives the admissible approximation of quantum evolution. 

The main requirements to the simulation of quantum dynamics through the collective behavior are the following.   
\begin{itemize}
\item Quantum dynamics is simulated by the dynamics of the swarm of samples so that in each time instant $t$ the quantum probability equals the density of the swarm.   
\begin{equation}
|\Psi(x,t)|^2=\rho(x,t)
\label{swarm}
\end{equation}
in each point $x$ of the configuration space. 
\item Each sample of the swarm has its own history, e.g. it preserves its individual number in course of the whole simulating process. The types of the samples exactly correspond to the types of real physical particles. 
\item The behavior of each sample is completely determined by its own state and the state of all samples in its close vicinity. 
\end{itemize}

The swarm satisfying these conditions is called the quantum swarm for one quantum particle. 

We define the behavior of the samples such that these conditions are satisfied. For this it is sufficient to show that for the solution $\Psi(x,t)$ of Shroedinger equation it is possible to move the samples only locally, e.g. to the small distance for the insurance of the equation (\ref{swarm}) in each time instant. Such a movement, of course, will be a priori non natural in the dynamical sense, but we will show how it can be done by means of the dynamical diffusion mechanism. 

 We note that the second rule means that we refuse from the using of complex numbers in the description of quantum mechanics. Also the locality of all interactions allows including QED to our model. The behavior of samples is the rule determining the change of its internal states (the type, impulse, momentum of impulse) and the spatial position (spatial shift). In view of the above mentioned the behavior cannot be determined by the classical physics. 

We define the quasi classical behavior of the samples called the dynamical diffusion mechanism. The swarm of samples with such behavior satisfies these conditions. 

Let us agree that each sample in each time instant can either to stay in place or to move along one of the coordinate axes $OX,OY,OZ$ with the speed $c$. 

We call the reaction of change the sequence of the following operations on the swarm:
the choice of pair $\a, \b$ of the samples located not farer than $\d x$ from each other, which speeds are mutually opposite: $v(\a )=-v(\b )$ and either simultaneous replacement of their speeds to zero (if they are nonzero) or acquiring them mutually opposite speeds of the module $c$ oriented along one of the coordinate axe. The axe is always chosen randomly from the uniform probability distribution. 

The reaction of change does not change neither summed impulse of the swarm, nor summed momentum of impulse if $\d x$ is sufficiently small. 
By $N(r)$ and $N_s(r)$ we denote the sets of all samples in the cube with the point $r$ and the set of samples from this cube possessing zero speeds correspondingly; by $N^+_x(r), N^+_y(r), N^+_z(r)$ we denote the sets of samples from the cube $r$, moving along the corresponding axe in the positive direction, and by analogous symbols with the sign $-$ the corresponding sets but for the negative direction of movement. By $|g|$ we denote the total number of samples in the set $g$. Let us agree to denote the total numbers of samples in a set by the same symbol as this set but with the replacement  $N$ to $n$. We call $r-$ stationary each subset $S\subseteq n(r)$ consisting of the samples with nonzero speeds for which $\sum_{\a\in S}v(\a )=0$ and $S$ is the maximal on including with this property. The total number $|S|$ of the samples of $r-$ stationary set (which does not depend on its choice) is denoted by $s(r)$. Let $d>0$ be a chosen constant such that the coefficient of the diffusion is proportional to $d$, $V(r)$ is a scalar field proportional to the external potential energy with the constant coefficient of proportionality, $grad\ V(r)=(V_x(r), V_y(r), V_z(r))$.  

We also agree to consider only non relativistic swarms, e.g. such that $n_s(r)/n(r)$ is close to $1$ for all $r$. It means that the bulk of samples in each cube have zero speed. This requirement is incompatible with the point wise approximation by the swarms \ref{asympt} of the exact wave functions for the external potential of Coulomb form $1/r$ because the mean speed of samples in the vicinity of zero point for such potentials must converge to infinity. For the asymptotic convergence \ref{asympt} we would have to assume that $c$ can be chosen as large as needed for every next swarm number $n$. In the reality $c$ cannot exceed the speed of light that establishes the natural limitation to the accuracy of the swarm approximation of the solutions of Shroedinger equation. 

The dynamical diffusion mechanism of evolution is the following sequence of the operation on the swarm:

\begin{itemize}
\item 1) The sequence of random reactions of change with the uniform distribution of probability, which gives the distribution of the speeds with the property $s(r)/n_s(r)=d$ for each point $r$. If $n(r)$ is small, this equality must be satisfied with the maximal accuracy (see agreement about the accuracy from above).
\item 2) The acquiring of speeds to some samples from $N_s(r)$, randomly chosen from the uniform distribution so that for each axe the signs of newly acquired speeds along this axe are the same, and if $v_u(r)$ is the summed speed acquired to the samples from $r$-th cube along the axe $u$, $u=x,y,z$, then for all such $u$ the equation $v_u(r)m=-V_u(r)$ is fulfilled with the maximal possible accuracy. 
\item 3) The change of coordinates $r(\a )$ of each sample corresponding to the Galileo law:
$r_{new}(\a )=r(\a )+v(\a )\Delta t$.
\item 4) Recalculation of $V(r)$ accordingly to the new positions of samples.
\end{itemize}

We do not concretize the method of recalculation of the potential energy. It may be done by Coulomb formula or by the diffusion mechanism as was proposed in the work \cite{Oz1}. 

The dynamical diffusion swarm cannot be represented as an ensemble of point wise particles with the classical interaction. The item 1) says about two things: 
\begin{itemize}
\item there is the force with the random direction which acts to the samples (compare with \cite{FH}), and
\item the samples acquire the mean value of speed within the accuracy determined by $\d x$ (the lesser $\d x$ is the more accurate mean value is taken).
\end{itemize}

For each time instant $t$ if $\Delta x$ is sufficiently small then the density of swarm $\rho (r,t)$ for any point $r$ will not depend on the orientation of the coordinate axes. 
Indeed, let $\d_1$ be such that $c\d t\ll\d_1 x \ll \d x$, and let $v(r)$ denote the average speed of the samples in the point $r$, found by the averaging on the samples with the coordinates $r_1:\ \| r-r_1\| < \d_1 x$. The total number of the samples occurred in the unit of time from the vicinity of the point $r_1$ to the vicinity of the close point $r_2$ will be then proportional to the dot product $v(r_1) (r_2-r_1)/\| r_2-r_1\|^2$, which does not depend on the orientation of the coordinate axes. 

A state of the dynamical diffusion swarm is determined by the coordinates and speeds of all its samples. This is its principal difference from the ensemble of DMC method where there are no speeds of the samples. 

\section{Differential equations for the dynamical diffusion swarm}

There is no system of differential equations on the density of the diffusion swarm and its average speed which is equivalent to Shroedinger equation. 

Nevertheless, there is the sequence of such systems which realizes the admissible asymptotic approximation of the solution of Shroedinger equation. Each of the system from this sequence depends on the fixed elementary length $\d x$. For example, the intensiveness of the diffusion process will beb proportional to $(\d x)^{-3}$. This does not allow to launch $\d x$ to zero as it is always done in the mathematical analysis when it is applied to the processes of classical physics. The value of grain $\d x$ must be chosen such that the approximation of the density $\rho = |\Psi |^2$ of the wave function by the density of the diffusion swarm within $(\d x)^3$ is satisfactory for the considered process. Only after the fixation of $\d x$ it is possible to build the diffusion swarm of the corresponding intensiveness and the differential equations approximating its dynamics which will be equivalent to Shroedinger equation.  

A state of the dynamical diffusion swarm is determined by the pair of functions 
\begin{equation}
\rho(t,\bar r),\ \bar p(t,\bar r),
\label{pair}
\end{equation}
where $\rho$ is the scalar function of density of the samples, $\bar p(\bar r)$ is the vector function resulting impulse of the samples in the point $\bar r$ defined as $\lim\limits_{dx\ar 0}P(r,dx)/(dx)^3$, where $P$ is the summed impulse of the samples occurred in the cube with $r$ with the side $dx$. Here we assumed that $dx$ can be done sufficiently less than the grain $\d x$, determining the coefficients of the equation on $\rho$ and $\bar p$. 

The dependence of equations on the grain $\d x$ will be revealed as follows. The summed impulse $\bar p(t,\bar r)$ varies slowly when $\bar r$ changes on values larger than $\d x$. But its derivative $\frac{\partial \bar p}{\partial t}$ will be very large: of the order $1/(\d x)^3$, and will vary rapidly as well. E.g., the graph of the function $\bar p(t,\bar r)$ is sufficiently smooth if we look at it with the large grain $\d x$, but if we raise the resolution by decreasing the grain $\d x$, we see that the graph looks like a saw with acute teeth. The more is resolution $1/\d x$ the more is sharpening of the teeth, it is limited by the limit of speed $c$ (compare with \cite{FH}). This important for further condition we call the non relativistic approximation and will write it as $v\ll c$. 

In view of isotropy of the diffusion process the change of density $\rho(r,t)$ in the time and its second derivative can be found by the integration on the surface of the sphere $S(r)$ of radius $\d x$ by formulas
\begin{equation}
\begin{array}{lll}
&\frac{\partial\rho(r,t)}{\partial t} &=\int\limits_{S(r)}\bar p(r,t)\bar n(\bar r_1)ds(r_1),\\
&\frac{\partial^2\rho(r,t)}{\partial t^2} &=\int\limits_{S(r)}\frac{\partial\bar p(r,t)}{\partial t}  \bar n(\bar r_1)ds(r_1).
\end{array}
\label{density}
\end{equation}
 These formulas are right for any mechanism of changing of the speed of samples. 

Now we derive the law of changing of the resulting impulse $\frac{\partial\bar p}{\partial t}\bar a$ of the swarm in the small sphere with center in the point $\bar r$, which results from the moveme of the samples along the vector $\bar a$ of normal to the surface of sphere of the unit length. 
Three magnitudes make the deposits to the change of the resulting impulse:
\begin{itemize}
\item Penetration of the samples which have acquired the speed in the reaction of change through the small element of the surface (diffusion). 
\item Penetration of the samples which have acquired the speed from the action of external potential.
\item Penetration of the samples which have preserved their speed (by inertia).
\end{itemize}
It follows from the definition of the diffusion process that these deposits equal correspondingly
$-I\ grad\ \rho\bar a$, $-\kappa\rho\ grad\ V\bar a$ and $g\rho\bar p\bar a$, where $I,\kappa,g$ is the intensities of the corresponding processes. The choice of units system allows us to make $g=1$. The dependence of the grain of spatial resolution needed for approximation of Shroedinger equation has the form:
\begin{equation}
I=\frac{h^2}{2m^2(\d x)^3},\ \kappa = \frac{h}{m\d x}.
\label{intens}
\end{equation}

In view of the non relativistic supposition we can omit the last summand which is sufficiently smaller than the first two for the small $\d x$.
We then obtain the following approximate formula.

\begin{equation}
\frac{\partial \bar p}{\partial t}\approx -I\ grad\ \rho-\kappa\rho\ grad\ V.
\end{equation}

The resulting equation on the density of the diffusion swarm has thus the form:
\begin{equation}
\frac{\partial^2\rho (r)}{\partial t^2}=-\int\limits_{S(r)}I\ grad\ \rho -\kappa\rho\ grad\ V)\bar n(r')\ dS(r'),
\label{swar}
\end{equation}
where the coefficients $I,\kappa$ can be found by \ref{intens}. 

We prove that the quantum swarm satisfies \ref{swar}, which means the admissible approximation of the quantum dynamics by the evolution of the swarm. 

The method of collective behavior permits to give the simple algorithm for the computation of the energy, impulse and momentum of impulse of quantum particle represented as the swarm of samples. Here accordingly to the quantum rules for the finding of any magnitude we have to averaging on the values of the reciprocal magnitude. To find the impulse we must fix some value of the time interval $\Delta t$ and fulfill averaging of impulses on all samples on any values of the distances they overcome. This receipt in our notations gives the vector
\begin{equation}
(\sum\limits_r cm\frac{n_x(r)^+-n_x(r)^-}{n(r)}, cm\frac{n_y(r)^+-n_y(r)^-}{n(r)}, cm\frac{n_z(r)^+-n_z(r)^-}{n(r)})
\end{equation}
which equals to the average impulse of all samples in the swarm found by averaging on the passes in the fixed time interval. The analogous calculation of the momentum of impulse or the potential energy gives the average momentum or the average potential energy.

When calculating the average kinetic energy we have to fix the pass $\d x$ and to take the average energy on all instant of the time $t$, because the time is the reciprocal magnitude for the energy. It means that we must sum only the energies of the samples which are moving in the considered time instant, e.g., for any cube for the speed $v=cn_x/n$ along the axe $x$ the expected total number of such samples is $nv/c$, and their fraction in the total number of samples is $n_x/n$. Here we use the non relativistic assumption that the total fraction of the moving samples is small. The total kinetic energy found by this rule is 
$$
mc^2n_x n_x/n+mc^2n_y n_y/n+mc^2n_z n_z/n
$$
which coincides to the kinetic energy found by the conventional formula $Mv_{mean}^2/2=((cn_x/n)^2+(cn_y/n)^2+(cn_z/n)^2)nm/2$. The laws of conservations for the impulse, momentum of impulse, energy of the swarm then follow from the classical laws of conservation and the non relativistic assumption. In the next paragraph we prove that the diffusion swarm dynamics can give the admissible approximation of the quantum one. Using Erenfest theorems and the laws of conservation for quantum dynamics we conclude that our method of calculation of these magnitudes $A$ gives their quantum mean values found by the formula $\langle A\rangle = \int\Psi^*(r)A\Psi(r)dr$. 

\section{About the diffusion swarm with non uniform intensity}

The intensiveness of the diffusion equals the coefficient of the Laplace operator in the diffusion equation. The intensiveness determines the total number of samples passing through the unit of square in the unit of time. To simulate quantum dynamics we need the diffusion swarm with the non uniform intensiveness. It means that the intensiveness of diffusion depends on the chosen grain of the spatial resolution $\d x$.
In this section we discuss how non uniform intensiveness of diffusion can be obtained in the same swarm. The concrete mechanisms discussed here will not be used further; they can be interesting only for the programming realization of the dynamical diffusion method. 

We at first consider the case when the external potential is constant $grad\ V=0$.
The dynamical diffusion process with the non uniform intensiveness can be insured by the special mechanism which we call threads. We illustrate the method on the following example. We suppose that all samples move not in all the space but along one closed trajectory (thread) which is determined by the smooth embedment of the circle to the space: $\g :\ S^1\ar R^3$. We suppose that the change of speeds happens only along this trajectory so that the samples remain in this thread in each time instant. This is equivalent to the imposing of the holonomic tie to the samples. We then suppose that the linear density as well as the module of speed of the samples is almost equal in all point of the trajectory. We consider the cube containing one point of this trajectory. The flow of samples through its border will not depend on its size $\d x$, because the thread is only one. The intensiveness of this process is thus proportional $1/(\d x)^3$, because the quantity of samples penetrating in unit of time into the cube with the side $\d x$ does not depend on $\d x$, and the density is obtained by the division of the quantity of samples to the volume. This example is not very good because many areas of the space remain without the samples at all. 

 We consider the next example. Let the space be divided to the cubes which are grouped by the layers $1,2,\ldots$. For each $j=1,2,\ldots$ the cubes of the layer $j+1$ consist of 8 cubes of the layer $j$, and their side, correspondingly, is twice large. For each $j$ the change of samples between the neighbor cubes of the layer $j$, occurring to the same cube of the layer $j+1$, goes only through the narrow channel with small capacity independent of $j$. The quantity of samples moving between the cubes of any fixed layer will not then depend on the number of this layer. It can be guaranteed by the appropriate choice of the pair for the change of impulses. Such a mechanism gives us the required intensiveness of the diffusion proportional to $1/(\d x)^3$, in view of the definition of density (\ref{density}). 

Now we consider the case of varying external potential. At each step of the evolution for the samples acquiring or loosing their speeds in the change we will use the rule from above, which insures the intensiveness proportional to $1/(\d x)^3$. The samples acquiring their speed from the action of the external potential will move as usual, independently of layers. We then obtain the formula (\ref{intens}).
This space design show how in principle the necessary non uniform intensiveness can be obtained in on swarm.

Practical realization of these methods presumes that we trace for the spatial location of the separated parts of the swarm, which means the refusal from the uniformity of the space and passage to the fractal space. The space with fractal dimensionality arises if we use the non uniform grid for the method of finite elements. This is why the proposed computational receipt says that we must fix the grain $\d x$ of the linear resolution such that the corresponding approximation of the wave function is satisfactory for our aims, and then consider the diffusion swarm with the intensiveness found by \ref{intens}. If we are not satisfied with the obtained dynamical picture, we must choose the new value of $\d x$ and repeat all the work. 

\section{Equivalence of quantum and dynamical diffusion swarms}

Here we show that the sequence of diffusion swarms represents the admissible approximation of quantum evolution. We have defined the quantum swarm as the swarm satisfying \ref{swarm} and which evolution can be represented as the local movements of the samples.  

At first we determine that the quantum swarm exists, e.g., that the equation \ref{swarm} can be really reached by only local shifts of the samples. Then we prove that the mechanism of the movements of samples coincides with the diffusion that gives the main result. 

Let us consider the quantum swarm. We start with Shroedinger equation
\begin{equation}
ih\frac{\partial\Psi(r,t)}{\partial t}=-\frac{h^2}{2M}\Delta\Psi(r,t)+V_{pot}(r,t)\Psi(r,t),
\label{Sh}
\end{equation}
which can be rewritten as 
\begin{equation}
\begin{array}{lll}
&\Psi^r_t(r)&=-\frac{h}{2M}\Delta\Psi^i_t(r)+\frac{V_{pot}}{h}\Psi^i(r),\\
&\Psi^i_t(r)&=\frac{h}{2M}\Delta\Psi^r_t(r)-\frac{V_{pot}}{h}\Psi^r(r)
\end{array}
\label{Sh2}
\end{equation}
for the real and imaginary parts $\Psi^r,\ \Psi^i$ of the wave function $\Psi$. We focus on the evolution of the density of quantum swarm, which is the function
$$
\rho(r,t)=(\Psi^r(r,t))^2+(\Psi^i(r,t))^2.
$$

Fixing the value of $\d x$ we apply for the approximation of the second derivative the difference scheme of the form
$$
\frac{\partial^2 \Psi(x)}{\partial x^2}\approx\frac{\Psi(x+\d x)+\Psi(x-\d x)-2\Psi(x)  }{(\d x)^2}
$$
for each time instant, where the wave function is supposed to satisfy all the sufficient conditions for such approximation. Since the addition of any constant to the potential energy $V_{pot}$ does not influence to the quantum evolution of the density, we can consider instead of $V_{pot}$ the other potential $V=V_{pot}+\a$, where $\a=-\frac{3h^2}{m(\d x)^2}$, that leads to the disappearing of the summand $2\Psi(x)$ in the difference schemes for the second derivative on $x,y,z$ (from which the coefficient $3$ arises) after its substitution to Shroedinger equation. We introduce the simplifying coefficient 
$$
\g = \frac{h}{2M}\frac{1}{(\d x)^2}.
$$
Since we yet do not know the mechanism of moving of the samples in quantum swarm, we suppose that we simply either take off some quantity of the samples from any cube or put it there from some storage. We split the evolution of the quantum swarm in the time to so small segments of the longitude  $\d t$, that in each of which the samples move only in the framework of two neighbor cubes. If we prove that the evolution of quantum swarm on such segment is insured by the diffusion mechanism, it will be right for the whole evolution as well. That is our supposition does not limit the generality. 
We also agree that these cubes differ from one another by the shift to $\d x$ along the axe $x$, that does not limit the generality as well. Let the centers of these cubes be $x$ and $x_1=x+\d x$. In these suppositions the summand $\Psi(x-\d x)$ in the difference scheme disappears as well and on the considered small time segment the evolution of quantum swarm is determined by the system of equations:
\begin{equation}
\begin{array}{lll}
&\Psi^r_t(x)&=-\g\Psi^i(x_1)+V(x)\Psi^i(x),\\
&\Psi^i_t(x)&=\g\Psi^r(x_1)-V(x)\Psi^r(x),
\end{array}
\label{quant}
\end{equation}
and by the analogous system obtained by the substitution of $x$ in place of $x_1$ and vise versa.

For such a segment we thus have  
\begin{equation}
\begin{array}{lll}
\frac{\partial\rho(x)}{\partial t}&=2\Psi^i(x)(\g\Psi^r(x_1)-V(x)\Psi^r(x))&+2\Psi^r(x)(-\g\Psi^i(x_1)-V(x)\Psi^i(x))=\\
&=2\g (\Psi^i(x)\Psi^r(x_1)-\Psi^r(x)\Psi^i(x_1))&=-\frac{\partial\rho(x_1)}{\partial t}.
\end{array}
\label{1der}
\end{equation}
It results in that the outcome of the samples in one cube equals their income to the other. The evolution of quantum swarm then satisfies the condition of locality. In the order to compare these evolution with the diffusion we now find the second derivative of the quantum density to time:
\begin{equation}
\begin{array}{ll}
&\frac{\partial^2\rho(x)}{\partial t^2}=2\g [ (\g\Psi^r(x_1)-V(x)\Psi^r(x))\Psi^r(x_1)+\Psi^i(x)(-\g\Psi^i(x)+V(x_1)\Psi^i(x_1))-\\
&(-\g\Psi^i(x_1)+V(x)\Psi^i(x))\Psi^i(x_1)-\Psi^r(x)(\g\Psi^r(x)-V(x_1)\Psi^r(x_1))]=\\
&2\g^2(\Psi^r(x_1))^2-2\g V(x)\Psi^r(x)\Psi^r(x_1)-2\g^2(\Psi^i(x))^2+\\
&2\g V(x_1)\Psi^i(x)\Psi^i(x_1)+2\g^2(\Psi^i(x_1))^2-2\g V(x)\Psi^i(x)\Psi^i(x_1)-\\
&2\g^2(\Psi^r(x))^2+2\g V(x_1)\Psi^r(x)\Psi^r(x_1)= \\
&2\g^2((\Psi^r(x_1))^2+(\Psi^i(x_1))^2-((\Psi^r(x))^2+(\Psi^i(x))^2))+\\
&2\g [(V(x_1)-V(x))((\Psi^r(x))^2+(\Psi^i(x))^2)+o(\d x)],
\end{array}
\end{equation}
where $o(\d x) = (\Psi^r(x)\Psi^r(x_1)+\Psi^i(x)\Psi^i(x_1)-((\Psi^r(x))^2+(\Psi^i(x))^2))(V(x_1)-V(x))$. Now we compare it with the expression for the second derivative of the density of diffusion swarm found in the previous section, taking into account that in our case the change of samples goes between two neighbor cubes along the axe $x$ only. Comparing with \ref{swar} in view of \ref{intens}, we conclude that the second derivative of the density of quantum swarm asymptotically converge to the second derivative of diffusion swarm. 

If we choose instead of the initial state the state where its density has Gauss form which is the ground state of harmonic oscillator, then for the corresponding value of energy $V=a(x^2+y^2+z^2)$ we have $\partial\rho /\partial t=0$ in the initial instant for any point of space. It is proved that the second derivative of the quantum swarm density and of the diffusion swarm density are the same, hence the diffusion swarm will be good approximation for the quantum density on some interval $\Delta T$. Switching on slowly some potential we will have also the approximation of any quantum evolution in the limit of swarms for the unlimited increasing of $n$. 

The swarm approximating quantum dynamics of one particle depends on the choice of $\d x$. After the fixation of $\d x$ we obtain for unlimitedly decreasing $\d t$ the approximation of the wave function within $\d x$. The intensiveness of the diffusion will be determined by $\d x$, it will be $\frac{h^3}{m^3c(\d x)^3}$. If we want to decrease the grain $\d x$ we must allow the more quantity of moving samples in the unit volume. It is necessary due to the uncertainty principle for length and impulse: the dispersion of speeds of the samples will grow if $\d x$ decreases. In any case for the obtaining of the dynamical picture one has to fix the grain of spatial resolution $\d x$. 

If the total number of samples $n$ is limited we obtain the model of quantum dynamics with decoherence of the inbuilt type. This model can be extended to the multi particle case where it can serve as the approximation of quantum dynamics in the standard Hilbert formalism (see (\cite{Oz1}). The robustness of this scheme for the numerical computations follows from that it gives Born rule for the quantum probability which thus turns to be inbuilt into the algorithmic formalism, in contrast to Copenhagen formalism where this rule is postulated. 

\section{Restoration of wave function from dynamical diffusion swarm}

We have solved the problem of approximation of the dynamics of density for one quantum particle by the special diffusion process with the non uniform intensity. A state of the dynamical diffusion swarm is determined by a pair \ref{pair}. Such a pair does not contain the notion of complex numbers which induced the famous quantum interference in the standard formalism. Furthermore, the diffusion swarm gives no beautiful differential equations of Shroedinger type for $\rho$ and $\bar p$. It radically differs from the classical processes (for example, heat transform or oscillations) because its intensity depends on the chosen grain of spatial resolution. We agree to these for the sake of the main: the economy of the computational resources required for the description of quantum dynamics. 

Now, to finalize the picture we have to solve the inverse problem: to show how to restore the wave function $\Psi$ from the given state of the dynamical diffusion swarm \ref{pair}. 
to do this we turn to the equality \ref{1der}, and substitute to it the expression of the wave function through the density:
 $\Psi (r)=\sqrt{\rho (r)}\exp(i\phi (r))$. We must find the phase $\phi (r)$ of the wave function. Since only relative phase has the physical sense we can fix some point $r$ and consider the phase of the other point $r_1$ relatively to $r$. If $r_1$ is close to $r$, the equation \ref{1der} gives us 
$$
\phi(r)-\phi(r_1)=arcsin\ k(\d x)^2\frac{\bar p(\bar r-\bar r_1)}{\sqrt{\rho(r)\rho(r_1)}}
$$
which results in the following formula for the finding of the relative phase:
\begin{equation}
\phi(r_1)=\int\limits_\g k(\d x)^2\bar v\ d\bar\g
\label{phase}
\end{equation}
where the contour $\g$ goes from $r$ to $r_1$. This definition depends explicitly on the choice of contour $\g$, hence we have to prove its correctness, e.g., its independence of the contour $\g$. Since the phase is determined within an integer multiplier of $2\pi$, the different choices of the contour can result at most in the change of the phase on such a number that takes place in case of excited states of electron in hydrogen atom with the nonzero momentum (for example, $3d$). We show that the integration of speed $\bar v$ of the swarm on the closed contour preserves its value in the time with the more exactness the less $\d x$ is. It involves that if in the initial time instant the definition (\ref{phase}) was correct, it then remains correct for the next time instants. 

We now consider the derivative of the integral of the speed for the closed contour $\g_c$. Applying the formula (\ref{swar}) and accounting $\partial\bar p /\partial t=\rho\ \partial\bar v/\partial t$, we obtain
\begin{equation}
\frac{\partial}{\partial t}\int\limits_{\g_c}(\d x)^2\bar v\ d\g=-\int\limits_{\g_c}I(\d x)^2\frac{grad\ \rho}{\rho}+\kappa (\d x)^2\ grad\ V.
\end{equation}
The first summand gives zero after the integration on the closed contour, because it is $grad\ \ln\rho$, the second summand gives zero by analogous reason.  

Now it is sufficient to verify the correctness of the definition (\ref{phase}) in the initial time instant that can be done immediately for any task. If the wave function of initial state can be obtained from the ground state in the Coulomb field where $\bar v=0$, then the correctness follows from the proved because there is no any phase shift here to $2\pi k$. If for the obtaining of the initial state in the considered problem we must start from some excited state with the shift of the phase, the correctness should be checked for this state at first. 

\section{Conclusion}

We have given the economical method of the simulation of one quantum particle dynamics. This method allows the creation of the video films showing the evolution of its state in the time. Our method is based on the mechanism of diffusion of the samples of swarm consisting of the point wise particles which obey the laws of classical mechanics with only one additional property: they change by impulses. The change of impulses creates the mechanism of diffusion for this swarm. The intensity of diffusion depends on the chosen grain of spatial resolution: the less the grain is the more is the intensity. This is why such a swarm cannot be described by a beautiful system of differential equations. Its main advantage over the methods of matrix algebra is the strict economy of the computational resources required for the simulation and the similarity to the classical description of the dynamics that permits to build the visual pictures of the evolution. 

\section{Acknowlegdements}

The author thanks academician Kamil Valiev for the constructive criticism.

\end{document}